\documentclass[twocolumn,showpacs,showkeys,preprintnumbers]{revtex4-1}
\usepackage{amssymb}

\usepackage{amsfonts}

\usepackage{latexsym}

\usepackage{dcolumn}

\usepackage{amsmath}

\usepackage{graphicx}

\usepackage{psfrag,pstricks}

\begin{document}

\title{ Quantum dynamics of an optomechanical system in the presence of Kerr-down conversion nonlinearity}

\author{S.Shahidani $^{1}$ }
\email{sareh.shahidani@gmail.com}

\author{M. H. Naderi$^{1,2}$}
\author{ M. Soltanolkotabi$^{1,2}$}
\author{Sh. Barzanjeh$^{1,3}$}
\affiliation{$^{1}$ Department of Physics, Faculty of Science, University of Isfahan, Hezar Jerib, 81746-73441, Isfahan, Iran\\
$^{2}$ Quantum Optics Group, Department of Physics, Faculty of Science, University of Isfahan, Hezar Jerib, 81746-73441, Isfahan, Iran\\
$^{3}$School of Science and Technology, Physics Division,
Universita di Camerino, I-62032 Camerino (MC), Italy}
\date{\today}

\begin{abstract}
We study theoretically nonlinear effects arising from the presence of  a Kerr-down conversion nonlinear crystal inside  an optomechanical cavity. For this system we investigate the influences of the two nonlinearities, i.e., the Kerr nonlinearity and the parametric gain,  on the dynamics of the oscillating mirror, the intensity and the squeezing spectra of  the  transmitted field, and the steady-state mirror-field   entanglement. We show that in comparison with a bare optomechanical cavity, the combination of the cavity energy shift due to the Kerr nonlinearity and increase in the intracavity photon number due to the gain medium can increase the normal mode splitting in the displacement spectrum of the oscillating mirror and reduce its effective temperature. Our work   demonstrates that both the Kerr nonlinearity and down conversion process increase the degree of squeezing in the transmitted field. Moreover, we find that in the system under consideration  the degree of entanglement between the mechanical and optical modes decreases considerably because of  the intracavity photon number reduction in the presence of the Kerr medium. 
\end{abstract} 

\pacs{37.30.+i, 03.67.Bg, 42.50.Wk, 42.50.Pq} 

\maketitle

\section{Introduction}
%
%

In recent years, there has been an increasing interest in cavity optomechanical systems  for a wide range of both experimental and theoretical investigations\cite{arci,corbitt1,schlei,vit,kipp,Groblacher,Xuereb,mancini1}.The importance of  optomechanical systems is due to their  potential applications  in different topics of physics.They  are promising candidates for studying quantum effects in the mesoscopic and macroscopic  scales\cite{marshal}, detection and interferometry of gravitational waves \cite{braginsky, corbitt, courty}and measurement of small displacements\cite{rugar}.
In a cavity optomechanical system the radiation pressure exerted by the electromagnetic field induces a coupling between the intensity of the cavity field and the mechanical motion of a movable mirror. It is pointed out that the necessary condition for observing  quantum phenomena for the oscillating mirror, as a mesoscopic or macroscopic object, is the  preparation of the mechanical oscillator at low phonon occupancy . Although the ground state cooling of the mirror has not yet been achieved experimentally,  it has been shown theoretically\cite{schlei,wilson,marq} that the ground state cooling  of the mirror is possible in the resolved sideband regime where the cavity bandwith is less than mechanical oscillation frequency of the mirror.
Another quantum phenomenon expected to be observed in  optomechanical sytems is the mixing between the fluctuations of the cavity field around the steady state and the mechanical mode which leads to the normal mode splitting(NMS) in the displacement spectrum of the  mirror\cite{dob}.  It is well known from cavity QED that  NMS is an absolute evidence  of   strong coupling between the subsytems with energy exchange taking place on a time scale faster than  the decoherence  of  each mode. The appearance of NMS in  optomechanical systems is known as  a direct consequence of ground state cooling of the mirror\cite{dob}.

One of the most important characteristics of optomechanical systems is their nonlinear optical peroperties. The strong interaction between the cavity filed and the mechanical oscillations makes the cavity behave like a nonlinear medium since the length of the cavity depends upon the intensity of the field in analogous way to the optical length of a nonlinear material\cite{mancini1}.
It has been shown\cite{nori} that  within the
 Born-Openheimer approximation the back action of the oscillating mirror in one edge of an optical Fabry-Perot cavity,driven by a nearly-resonant laser field, induces an intracavity  Kerr-like nonlinearity.  
  Besides this intrinsic nonlinearity,  the presence of an optical parametric amplifier (OPA) \cite{huang} or  the optical Kerr medium \cite{kumar} inside the cavity   has opened up a new domain for combining nonlinear optics and optomechanics towards the enhancement of quantum effects.  It has been demonstrated\cite{huang}  that the presence of  an OPA in the cavity causes a strong coupling between the oscillating mirror and the cavity mode resulting from increasing  the intracavity photon number. Thus the enhancement of radiation pressure-induced coupling not only considerably improves the cooling of the mechanical oscillator but also makes the observation of the NMS of the movable mirror and the output field mode accesible\cite{huang}. On the other hand, when the optomechanical cavity contains an optical Kerr medium with $ \chi^{(3)} $ nonlinearity, due to the photon-photon repulsion and the reduction of the cavity photon fluctuations, the normal mode splitting weakens and the effective temperature of the moving mirror increases\cite{kumar}.

     The above mentiond  interesting   results motivated us towards investigation  of an optomechanical system which contains  a nonlinear crystal consisting of a Kerr medium and a degenerate OPA. We will show that in the system under consideration the competition between the increasing of intracavity photon number due to the parametric gain process and the reduction of the number of photons due to the photon blockade mechanism arising from the presence of the Kerr medium leads  to some new interesting effects in the dynamics of the movable mirror in the low photon number  limit and the  resolved sideband regime.

   The paper is structured as follows. In Sec.\ref{sec1} we describe the model, derive the quantum Langevin equations of motion for the system operators and find their steady-state mean values. In Sec.\ref{sec2} we linearize the quntum Langevin equations of motion  around the steady-state mean values, and analyze  the stability conditions of the system. In Sec.\ref{sec3} we calculate the spectrum  of small fluctuations in the position of the oscillating  mirror , the effective frequency and effective damping rate of the mechanical oscilltor, and analyze the influence of the nonlinear gain and anharmonicity parameter on them. In Sec.\ref{field} we investigate the influences of the parametric gain and Kerr nonlinearity on the  intensity and quadrature squeezing of the transmitted field. In Sec.\ref{ent} we examine the entanglement between the optical and the mechanical mode.    Finally, we summarize our conclusions in Sec.\ref{sum}.
   

\section{The Physical Model}\label{sec1}
We consider a Kerr-down conversion optomechanical system composed of a  degenerate OPA and a nonlinear Kerr medium placed within a Fabry-Perot cavity formed by a fixed partially transmitting mirror and one movable perfectly reflecting mirror  in equilibrium with a thermal bath at  temperature  $ T$. The movable mirror  is free to move along the cavity axis and is treated as a quantum mechanical harmonic oscillator with effective mass $m$, frequency $\omega_{m}$ and energy decay rate $\gamma_{m}=\omega_{m}/Q$(where $Q$ is the mechanical quality factor).The cavity field is coherently driven by an input laser field with frequency $\omega_{L}$ and amplitude $\epsilon$ through the fixed mirror. Furthermore, the system is pumped by a coupling field to produce parametric oscillation and induce the Kerr nonlinearity in the cavity. In our investigation, we restrict the model to the case of single-cavity and mechanical modes\cite{law, genes}. The single cavity-mode assumption is justified in the adiabatic limit, i.e., $\omega_{m}<<\pi c/L$ which $c$ is the speed of light in vacuum and $L$ is the cavity length in the absence of the cavity field. We also assume that the induced resonance frequency shift of the cavity and the nonlinear parameter of the Kerr medium are much smaller than the longitudinal-mode spacing in the cavity. Furthermore, one can restrict to a single mechanical mode when the detection bandwidth is chosen such that it includes only a single, isolated, mechanical resonance and mode-mode coupling is negligible. It should be noted that in the adiabatic limit, the number of photons  generated by the Casimir, retardation, and Doppler effects is negligible \cite{mancini, man-tomb,giov}.
\begin{figure}[ht]
\centering
\includegraphics[width=3.5 in]{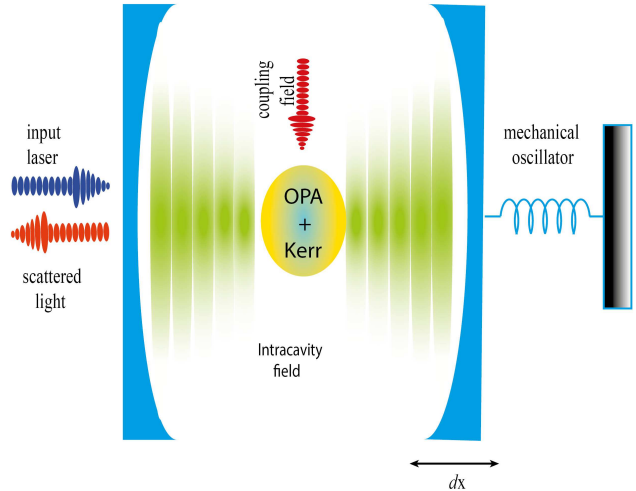}
\caption{
 (Color online) Schematic picture of the setup studied in the text. The cavity contains a Kerr-down conversion system which is pumped by a coupling field to produce parametric oscillation and induce Kerr nonlinearity in the cavity.}
\label{fig:fig1}
\end{figure}

Under these conditions, the total  Hamiltonian of the system  in a frame  rotating at the laser frquency $\omega_{L}$can be written as
\begin{equation} \label{H}
H=H_{0}+H_{1} 
 \end{equation}
where,
\begin{subequations}\label{hamil}
\begin{eqnarray}
H_{0}&=&\hbar(\omega_{0} -\omega_{L}  ) a^{\dagger} a+ ( \frac{ p^2} {2m} +\frac{1} {2} m \omega_{m}^2 q^2) -\hbar g_{m} a^{\dagger} a q\nonumber\\
&&+
i\hbar\epsilon(a^{\dagger}-a),\\  
H_{1}&=&i \hbar G( e^ { i\theta} a^{\dagger 2}  -e^ {-i\theta} a^2 )+\hbar\eta a^{\dagger 2} a^{2} .
\end{eqnarray}
\end{subequations} 
The first two terms in $H_{0} $ are,respectively, the free Hamiltonian of the cavity field with annihilation(creation) operator   $a(a^{\dagger})$  and the movable mirror with resonance frequency $\omega_{m}$ and effective mass $m$, the third term describes the  optomechanical coupling between the cavity field and the mechanical oscillator due to radiation pressure force,  and the last term in $H_{0} $  describes the driving of the intracavity mode with the input laser.  Also, the two terms in  $H_{1} $ describes, respectively, the coupling of the intracavity field with  the  OPA and the Kerr medium; $G$ is the nonlinear gain of the OPA which is proportional to the pump power driving amplitude, $\theta$ is the phase of the field driving the OPA, and $\eta$ is the anharmonicity parameter proportional to the  third order nonlinear susceptibility $\chi^{(3)}$  of the Kerr medium.
The input laser field populates the intracavity mode through the partially transmitting mirror, then the photons in the cavity will exert a radiation pressure force on the movable mirror.  In a realistic treatment of the dynamics of the system , the cavity-field damping due to the  photon-leakage through the incomplete mirror and the Brownian noise associated with the coupling of the oscillating mirror to its thermal bath should be considered. Using the input-output formalism of quantum optics\cite{gardiner}, we can consider the effects of these sources of noise and dissipation in the quantum Langevin equations of motion. For the given Hamiltonian (\ref{H}), we obtain the following nonlinear equations of motion
\begin{subequations}\label{langevin}
\begin{eqnarray}
\dot{q}&=& \dfrac{p}{m},\\
\dot {p}&=&-m \omega_{m}^{2} q +\hbar g_{m}a^{\dagger}a-\gamma_{m} p+\xi ,\\
\dot{a}&=&-i (\omega_{0}-\omega_{L})a+i g_{m} q a +\epsilon -2 i \eta a^{\dagger}a^{2}   \nonumber\\ 
 &&+
2G a^{\dagger}e^{i\theta} -\kappa a +\sqrt{2\kappa}a_{in},
\end{eqnarray}
\end{subequations}
where $\kappa$ is the cavity  decay rate through the input mirror and $a_{in}$ is the input vacuum noise operator that is  charactrized by the following correlation functions  \cite{gardiner}
\begin{subequations}\label{noise1}
\begin{eqnarray}
<\delta a_{in}(t)\delta a_{in}^{\dagger}(t')>&=& \delta(t-t'),\\
<\delta a_{in}(t)\delta a_{in} (t')>&=&<\delta a_{in}^{\dagger} (t)\delta a_{in}^{\dagger}(t')>=0 .
\end{eqnarray}
\end{subequations}
The Brownian noise operator $\xi$ describes the heating of the mirror by the thermal bath at temperature $T$ and is characterized by the following correlation function  \cite{giov}
\begin{equation}\label{noise2}
<\xi(t)\xi(t')>=\frac{\hbar \gamma_{m}m}{2\pi} \int \omega e^{-i\omega(t-t')}[\coth(\frac{\hbar\omega}{2k_{B} T} )+1]d\omega,
\end{equation}
where $k_{B} $ is the Boltzmann constant. 

We are interested in the  steady-state regime and in small fluctuations with respect to the steady state. Thus we obtain the steady-state mean values of $p$, $q$ and $a$ as
  \begin{equation}\label{qs}
 p_{s}=0, \hfill\hfill q_{s}=\frac{\hbar g_{m}}{m \omega_{m}^{2}}|a_{s}|^{2},
 \end{equation}
 \begin{equation}\label{as}
a_{s}=\frac{\kappa -i\Delta^{'} +2G e^{i\theta}}{\Delta ^{'2}+\kappa^{2}-4G^{2}}\epsilon ,
\end{equation}
where  $q_{s}$ denotes  the new equilibrium position of the movable mirror and $\Delta^{'}=\omega_{0}-\omega_{L}-g_{m} q_{s} +2\eta |a_{s}|^{2}=\Delta+2\eta |a_{s}|^{2} $ is the effective detuning of the cavity which includes both  the radiation pressure  and the  Kerr medium effects. It is obvious that the optical path and hence the cavity detuning are modified in an intensity-dependent way. The first modification which is a mechanical nonlinearity, arises from the  radiation pressure-induced coupling between the movable mirror and the cavity field and the second modification  comes from the presence of the nonlinear Kerr medium in the optomechanical system. It has   been shown that \cite{imam} for a  given  sufficiently large pure  $\chi^{(3)}$ nonlinearity
inside an optical cavity, the effect of photon blockade occurs as a consequence of large phase
shifts of the cavity detuning. It has also been  predicted \cite{rabl}  that the same  photon-photon interaction can occur in the strong coupling regime in the optomechnaical systems. In our  treatment the photon-photon interaction due to the  radiation pressue is ignorable in comparison with  photon-photon interaction due to the Kerr nonlinearity.  Since $\Delta^{'} $ satisfies a fifth-order equation, it  can have five real solutions and hence the system may exhibit multistability for a certain range of parameters. 
\section{Dynamics Of Small Fluctuations}\label{sec2}
In order to investigate the dynamics of the system, we need to study the dynamics of small fluctuations near the steady state. We assume that the nonlinearity in the system is weak and decompose  each operator in Eq.(\ref{langevin}) as the sum of its steady-state value and a small fluctuation with zero mean value,
\begin{equation}\label{linear}
a=a_{s}+\delta a,\,\,\, q=q_{s}+\delta q,\,\,\, p=p_{s}+\delta p.
\end{equation}
Inserting the above linearized forms of the system operators into Eq.(\ref{langevin}), the linearized quantum Langevin equations  for the  fluctuation operators take the form 
\begin{subequations} \label{fluc}
\begin{eqnarray}
\dfrac{d}{dt}\delta q&=& \delta p/m,\\
\dfrac{d}{dt}\delta p&=&-m \omega_{m}^{2} \delta q+\hbar g_{m} (a_{s}\delta a^{\dagger}+a_{s}^{*}\delta a )-\gamma_{m}\delta p+\xi, \\
\dfrac{d}{dt}\delta a&=&-(i \Delta+\kappa)\delta a -i g_{m} a_{s}^{*} \delta q +(2G e^{i\theta} -2i\eta a_{s}^{2} )\delta a^{\dagger} \nonumber\\
&&-
4 i\eta |a_{s}|^{2} \delta a+\sqrt{2\kappa}\delta a_{in},\\
\dfrac{d}{dt}\delta a^{\dagger}&=&(i \Delta-\kappa)\delta a^{\dagger}+i g_{m} a_{s}\delta q +(2G e^{-i\theta} +2i\eta a_{s}^{*2} )\delta a \nonumber\\
&&+
4 i\eta |a_{s}|^{2}  \delta a^{\dagger} +\sqrt{2\kappa}\delta a_{in}^{\dagger} .
\end{eqnarray}
\end{subequations}
 Defining the cavity field quadratures $\delta x=\delta a+\delta a^{\dagger} $ and $\delta y=i(\delta a^{\dagger}-\delta a) $ and the input noise quadratures $\delta x_{in}=\delta a_{in} +\delta a^{\dagger}_{in}  $ and $\delta y_{in} =i(\delta a^{\dagger}_{in} -\delta a_{in})  $  Eq.(\ref{fluc})  can be written in the compact matrix form
 \begin{equation}\label{matrixform}
 \dot{u}=M u(t)+n(t),
 \end{equation}
where $u(t) =(\delta q,\delta p,\delta x,\delta y)^{T}$ is the vector of fluctuations, $n(t) =(0,\xi ,\sqrt{2\kappa}\delta x_{in},\sqrt{2\kappa} \delta y_{in})^{T}$ is the vector of noise sources and the  matrix $M$ is given by
\begin{equation}
M=\left(\begin{array}{ccccccc}
0 & 1/m & 0&0  \\
-m\omega_{m}^{2} &-\gamma_{m} &\hbar a_{+}   &-i\hbar a_{-} \\
   2i a_{-}  & 0 & -\kappa+\Gamma_{1}& \Delta_{1}+\delta_{1}  \\
    2 a_{+}& 0 & -\Delta_{1}+\delta_{1}  & -\kappa-\Gamma_{1} \\
   \end{array}\right),
\label{M}
\end{equation}
where we have defined 
\begin{eqnarray}\label{define}
a_{\pm}&=& g_{m}(a_{s}\pm  a_{s}^{*})/2,\nonumber
\\
\Gamma_{1}&=&2G \cos \theta-i \eta (a_{s}^{2}-a_{s}^{*2}),\nonumber
\\
\Delta_{1}&=&\Delta +4\eta \vert a_{s}\vert^{2} \\
\delta_{1}&=& 2G \sin \theta - \eta (a_{s}^{2}+a_{s}^{*2}).
\end{eqnarray}
Here, we concentrate on the stationary properties of the system. For this purpose, we should consider the steady-state condition governed by Eq.(\ref{matrixform}). The system is stable only if the real part of all eigenvalues of the matrix $M$ are negative, which is also the requirement of the validity of the linearized method. 
The parameter region in which the system is stable can be obtained from the Routh-Hurwitz criterion\cite{routh}, which gives the following three independent conditions:
\begin{subequations}
\begin{eqnarray}
s_{1}&=&2 \kappa  R_{1} +\omega_{m}^{2}(\gamma_{m}+4\kappa^{2})>  0, \\
s_{2}&=&\lbrace R_{1} \omega_{m}^{2}- 4\frac{ h g_{m}^{2}}{m} \eta|a_{s}|^{4}
+
i G (a_{s}^{2}e^{- i\theta}
 \nonumber\\
&-&c.c.)\rbrace > 0,\\
s_{3}&=&-[R_{1} \omega_{m}^{2}- R_{2}]\times\omega_{m}^{2} (\gamma_{m}
+2\kappa)+ [2\kappa  R_{1}\nonumber\\
&+&
 \omega_{m}^{2}(\gamma_{m} +4\kappa^{2})]>  0,
\end{eqnarray} \label{routh1}
\end{subequations}
where we have defined $R_{1}$ and $R_{2}$ as
\begin{subequations}
\begin{eqnarray}
R_{1}&=&\Delta_{1}^{2}-4G^{2}+\kappa^{2}+12\eta^{2}|a_{s}|^{4}\nonumber\\ 
&&+
8\eta\Delta_{1}|a_{s}|^{2} +4iG \eta(a_{s}^{2}e^{- i\theta}-c.c.),\\
R_{2}&=&2\frac{ h g_{m}^{2}}{m}(2\eta |a_{s}|^{4}+
i G (a_{s}^{2}e^{-i\theta}-c.c.)).
\end{eqnarray}
\end{subequations}
 \section{Displacement Spectrum of the oscillating mirror}\label{sec3}
Since we are interested in the spectrum of fluctuations in position of the oscillating  mirror  in the presence of Kerr-down conversion nonlinearity,
it is convenient to solve the time-domain equation of motion (\ref{matrixform}) by Fourier transforming it into the frequency domain. The Fourier transform of the time-domain operator $u(t)$ is
\begin{equation}
\tilde{u}(\omega)=\dfrac{1}{2\pi}\int_{-\infty}^{\infty} dt e^{-i\omega t}u(t). 
\end{equation}
Then solving the linearized quantum Langevin equation for the position fluctuations of  the oscillating mirror  yields
\begin{equation}\label {deltaq}
\delta q(\omega)=F_{1}(\omega) \xi(\omega)+F_{2}(\omega)\delta a_{in}(\omega)+F_{3}(\omega)\delta a_{in}^{\dagger}(\omega) ,
\end{equation}
in  which 
\begin{eqnarray} \label{coefficient}
F_{1}(\omega)&=&\frac{1}{m d(\omega)}\lbrace(-\kappa-i\omega)^{2}+\beta _{2}\rbrace ,\nonumber\\
F_{2}(\omega)&=&\frac{\sqrt{2\kappa}\hbar g_{m} }{m d(\omega)} [-a_{s}^{*}(-\kappa+i\Delta^{'}-i\omega)+2G e^{-i\theta}a_{s}],\nonumber\\
F_{3}(\omega)&=&F_{2}(-\omega)^{*},
\end{eqnarray}
with
\begin{equation}\label{d}
d(\omega)=
(\omega _{m}^{2}-\omega^{2}+i\omega \gamma_{m})(\beta _{2}+(\kappa +i\omega)^{2})-\beta _{1},
\end{equation}
\begin{eqnarray} \label{beta}
\beta _{1}&=&\frac{2\hbar g_{m}^{2}}{m}[\Delta^{'} |a_{s}|^{2}+iG (a_{s}^{2}e^{-i\theta}-c.c.)],\nonumber\\
 \beta_{2}&=&-4G^{2}+\Delta_{1}^{2}-4\eta^{2}|a_{s}|^{4}+
4i G \eta
(a_{s}^{2}e^{-i\theta}-c.c.).\nonumber\\
 \end{eqnarray}
The first term in Eq.(\ref{deltaq}) originates from the thermal noise while the second and third terms arise from the  radiation pressure force . The displacement spectrum of the mirror is defined by
\begin{equation}
S_{q}(\omega) =\frac{1}{4\pi}\int d\Omega e^{-i(\omega+\Omega)t}<\delta q(\omega)\delta q(\Omega)+\delta q(\Omega)\delta q(\omega)>.
\end{equation}
Using the following correlation functions in the frequency domain
\begin{equation}
<\delta a_{in}(\omega)\delta a_{in}^{\dagger}(\Omega)>=2\pi\delta(\omega+\Omega),
\end{equation}
\begin{equation}
<\xi(\omega)\xi(\Omega)>=2\pi\hbar \gamma_{m}\omega[1+\coth(\frac{\hbar\omega}{2k_{B}T})]\delta(\omega+\Omega),
\end{equation} 
the displacement spectrum of  the oscillating mirror is  obtained as
\begin{eqnarray}\label{dis spectrum} 
S_{q}(\omega)&=&\hbar| \chi|^{2}\lbrace m\gamma_{m}\omega\coth(\frac{\hbar\omega}{2k_{B}T})+2\frac{(\hbar g_{m})^{2}\kappa}{m^{2}|d(\omega)|^{2}}\nonumber\\
&&\times
\frac{|a_{s}|^{2}\delta_{2}^{2}+4GRe(a_{s}^{2}e^{-i\theta}(\kappa+i\Delta^{'})) }{(\beta_{2}+\kappa^{2}-\omega^{2})^{2}+4\kappa^{2}\omega^{2}}\rbrace,
\end{eqnarray}
 where
 $\delta_{2}^{2}= (\kappa^{2}+\omega^{2} +4G^{2}+\Delta^{'2})$.
\subsection{Normal Mode Splitting} 
In order to determine the structure of the  displacement spectrum of the moving mirror we need to determine the eigenvalues of the matrix $iM$ as the solution of Eq.(\ref{matrixform}) in the frequency domain, or the  zeros of $d(\omega)$ in the denominator of coefficients in Eq.(\ref{coefficient}). First we assume that the radiation pressure coupling $g_{m}$ is zero and find the eigenvalues of $iM$ as
\begin{equation}
-\frac{i\gamma_{m}}{2}\pm \sqrt{\omega_{m}^{2}-\frac{\gamma_{m}^{2}}{4}},\,\,-i\kappa \pm\sqrt{\beta_{2}} .
\end{equation}
Thus in the absence of optomechanical coupling the effective frequency of the normal modes depends on $\beta_{2}$ which is a function of nonlinear coefficiens $\eta$ ,$G$ and intracavity intensity .  
Now we assume that  the optomechanical coupling $g_{m}$ is not zero, but the  decay rates $\gamma_{m}$ and $\kappa$ are negligibly small in comparison with $\omega_{m}$ and $\beta_{2}$ . Then  the zeros of  $d(\omega)$ are approximately given by
\begin{equation}\label{normal}
\omega_{\pm}^{2}\cong \frac{1}{2}(\omega_{m}^{2}+\beta_{2}\pm\sqrt{(\omega_{m}^{2}-\beta_{2})^{2}+4\beta_{1}}).
\end{equation}
Hence, the Kerr-down conversion  medium can alter  the   radiation pressure contribution to the  NMS in two ways. Firstly, both nonlinearity parameters $G$ and $\eta$ change  the mean intracavity photon number considerably. Secondly the    parameter  $\beta_{1}$ is a function of the gain and anharmonicity parameters. 
  
\begin{figure}[ht]
\centering
\includegraphics[width=3.5 in]{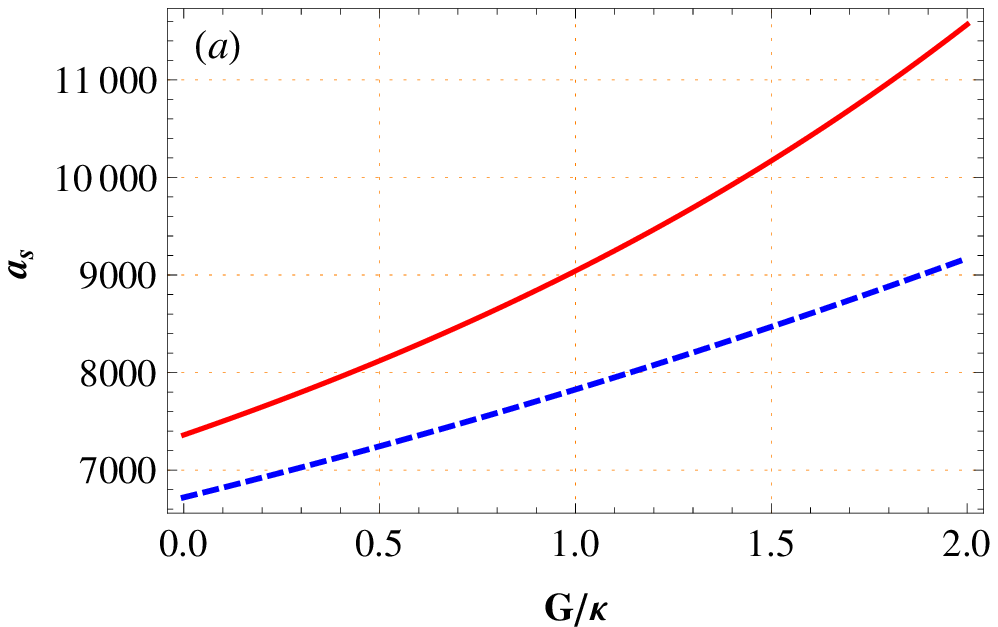}
\includegraphics[width=3.5 in]{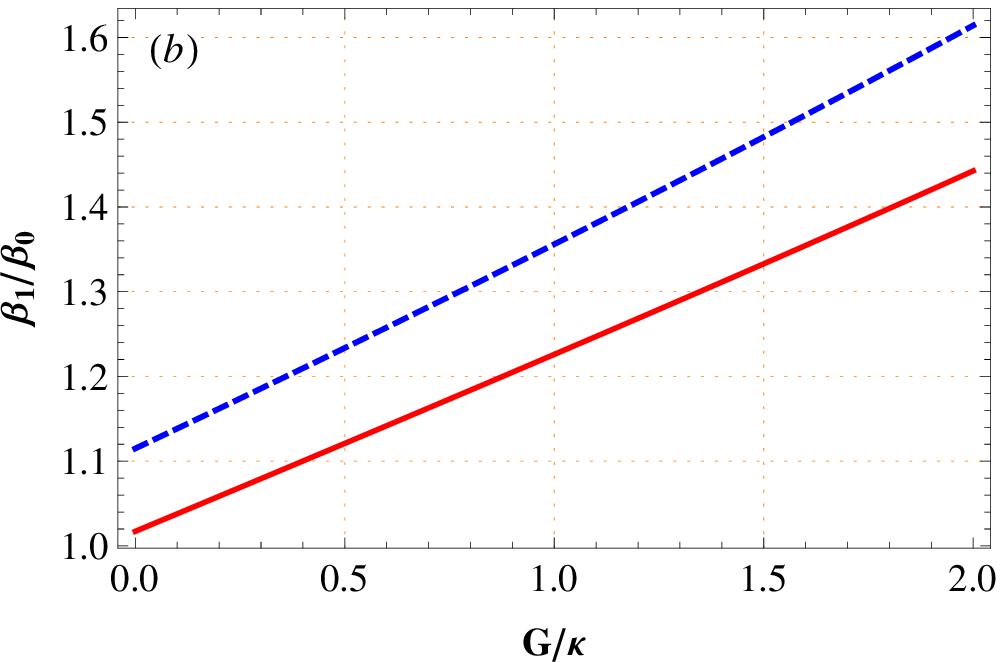}
\caption{
(Color online)(a)The mean intracavity photon number and (b) the normalized parameter $\beta_{1}/\beta_{0} $ versus the  normalized  gain $G/\kappa$  for $\eta=0.01$Hz (red solid line)  and  $\eta=0.05$ Hz (blue dashed line).The parameters are
 $\omega_{m}/2\pi=10$  MHz,$L=1$mm, $m=10$ng, $Q= 5\times10^5 $ Hz,   input laser power $P=6.9$ mW at $\lambda=1064$ nm,  $\kappa=0.1\omega_{m}$,  and $\theta=\pi/2$ .
}
\label{fig:beta}
\end{figure}

In Fig.\ref {fig:beta}, by using the experimental parameters from Ref.\cite{teufel},  we have plotted the mean intracavity photon number $ a_{s} $ and  the normalized parameter $\beta_{1}/\beta_{0}$  versus the normalized  gain $G/\kappa$ for   two different values of anharmonicity parameter $\eta$, where $\beta_{0}= 2\hbar g_{m}^{2}(\Delta |a_{s}|^{2}/m)$ is the contribution of  radiation pressure  to the normal modes of displacement spectrum in the bare cavity.  The figure shows  that while  there is a reduction in the number of intracavity photons (hence the radiation pressure)  due to the photon-photon repulsion mechanism (in the presence of Kerr medium), the contribution of radiation pressure coupling to the normal modes which is contained in $\beta_{1}$ can be increased by choosing proper values  of initial detuning and nonlinear coefficients $G$,    $\eta$ and $\theta$ in such a way that $\beta_{1}> \beta_{0}$ .
\begin{figure}[ht]
\centering
\includegraphics[width=3.5in]{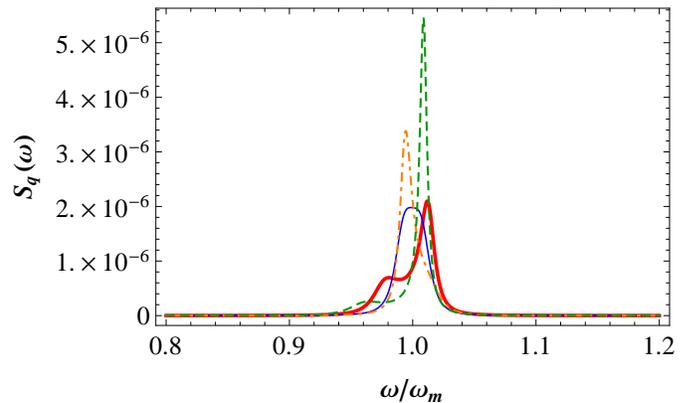}
\caption{
(Color online)The  displacement spectrum  versus the normalized frequency $\omega/\omega_{m}$ for bare cavity $(\eta=0,  G=0)$(blue thin curve), for cavity with Kerr nonlinearity $(\eta=0.01Hz, G=0)$ (orange dashed-dotted curve), with gain nonlinearity $(\eta=0 , G=8\times10^{6}Hz)$ (green dashed curve) and  with both nonliearity$(\eta=0.01Hz , G=8\times10^{6}Hz)$ (red thick curve). The parameters are $\omega_{m}/2\pi=10$  MHz,$L=3$mm,  $m=12$ng, $Q= 5\times10^5 $ Hz,   input laser power $P=9$ mW at $\lambda=1064$ nm,  $\kappa=0.02\omega_{m}$,  and $\theta=\pi/2$ . }
\label{fig:figc}
\end{figure}

\begin{figure}[ht]
\centering
\includegraphics[width=3.5in]{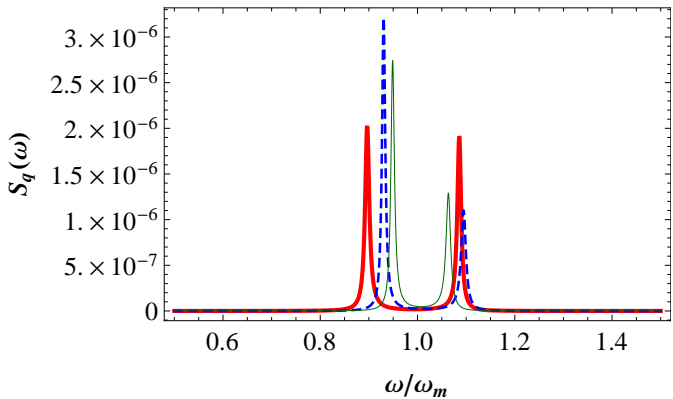}
\caption{
 (Color online)The  displacement spectrum versus the normalized frequency $\omega/\omega_{m}$ for  different values of  the parametric gain: $G=0$(green thin curve), $G=10^{5}$Hz(blue dashed curve), and $G=10^{7}$Hz(red thick curve). The parameters are $\eta=0.01$ Hz,  $\Delta=\omega_{m}$,  $\theta=\pi/2$,  $\omega_{m}/2\pi=10$  MHz,$L=1$mm, effective mass $m=10$ng, $Q= 5\times10^5 $ Hz, $P=30$ mW, $\lambda=1064$ nm, and $\kappa=0.01\omega_{m}$.}
\label{fig:dis-G}
\end{figure}

\begin{figure}[ht]
\centering
\includegraphics[width=3in]{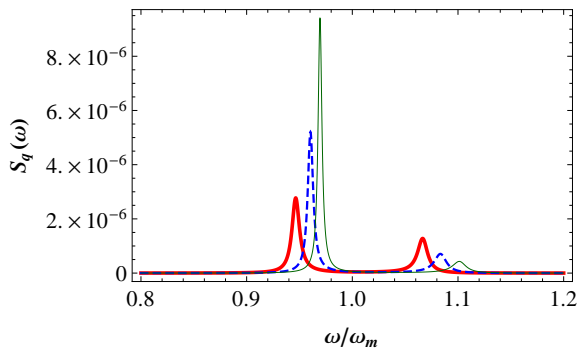}
\caption{
 (Color online)The  displacement spectrum  versus the normalized frequency $\omega/\omega_{m}$ for  different values of  the anharmonicity: $\eta=0.01$Hz(red thick curve) $\eta=0.03$Hz(blue dashed curve), and $\eta=0.06$Hz(green thin curve).The parameters are $G=10^6$Hz , $\Delta=\omega_{m}$,  $\theta=\pi/2$,  $\omega_{m}/2\pi=10$  MHz,$L=1$ mm, $m=10$ng, $Q= 5\times10^5 $ Hz,  $P=30$ mW, $\lambda=1064$nm, and $\kappa=0.01\omega_{m}$.}
\label{fig:dis-eta}
\end{figure}

 It is well known that the NMS in a bare-damped optomechanical system occurs only if  $g_{0}|a_{s}|\gtrsim \kappa$ (where  $g_{0}=\sqrt{\hbar/2m \omega_{m}}g_{m}$) due to the finite width of the peaks\cite{dob,marq}. In Fig.(\ref{fig:figc}) we have plotted the displacement spectrum of the movable mirror  below this threshold ($g_{0}|a_{s}|\gtrsim \kappa$ ). As expected,   the radiation pressure-induced coupling between the oscillating mirror and the fluctuations of the cavity mode does not lead to NMS. The presence of  the Kerr medium  alone dose not change this situation, while in the presence of the gain medium solely, mode splitting occurs only when the parametric gain $G$ is   large enough. However, as  shown in Fig.(\ref{fig:figc}), the combination of these two nonlinearities  leads to the appearance of NMS in the displacement spectrum. We can come to this conclusion  from the fact that in the peresence of the two nonlinearities, there is a competition between the two  quantum processes; the nonlinear gain medium increases the intracavity photon number and splits up the pump photons into two degenerate subharmonic photons , while  the Kerr medium shifts the energy levels of the cavity proportional to the anharmonicity parameter $\eta$ which blocks the entrance of a large number of photons into the cavity. The combination of the  energy shift resulting from the Kerr medium and increasing the photon number due to  the gain medium leads to the appearence of NMS. It is notable that  the presence of the gain medium  decreases the degree of photon-photon repulsion due to the Kerr nonlinearity, and  hence increases the parameter $\beta_{1}$  in Eq.(\ref{normal})and the radiation pressure effect on the oscillating mirror.
In Fig.(\ref{fig:dis-G}) and  Fig.(\ref{fig:dis-eta})  we have plotted the displacement spectrum of the movable mirror above the threshold of splitting ($g_{0}|a_{s}|\gtrsim \kappa$) for different values of the gain coefficient $G$ and different values of anharmonicity parameter  $\eta$, respectively. As expected, the presence of the gain medium inside the cavity strengthens the coupling between the oscillating mirror and the cavity mode, while the presence of the Kerr medium inside the cavity  weakens this coupling due to the photon-photon repulsion and shifts the modes far from each other. 

\subsection{Effective frequency and effective damping rate of the oscillating mirror }
\begin{figure} [ht]
\centering
\includegraphics[width=3.5 in]{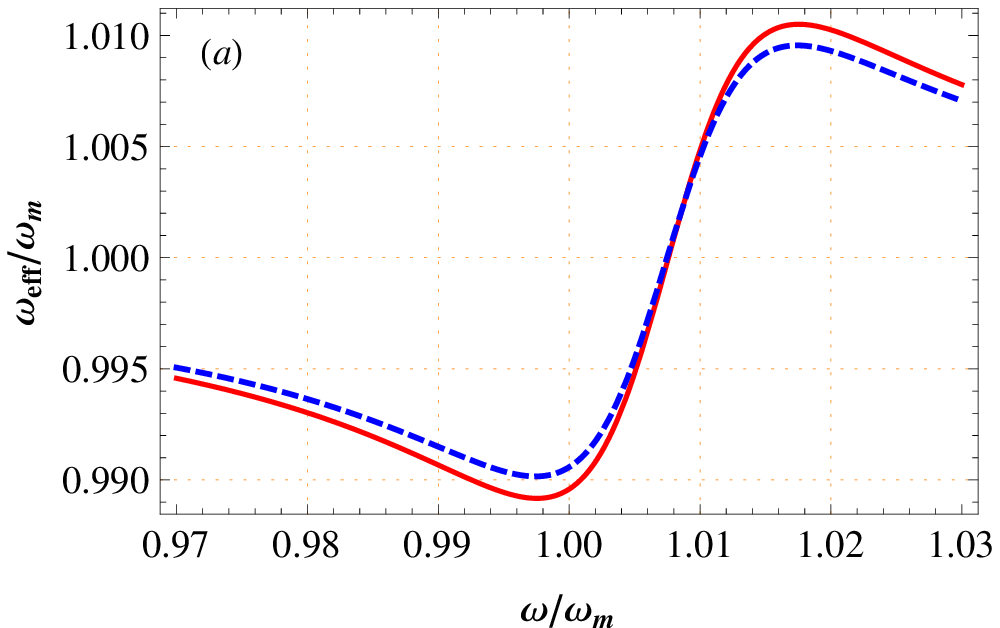}
\includegraphics[width=3.5 in]{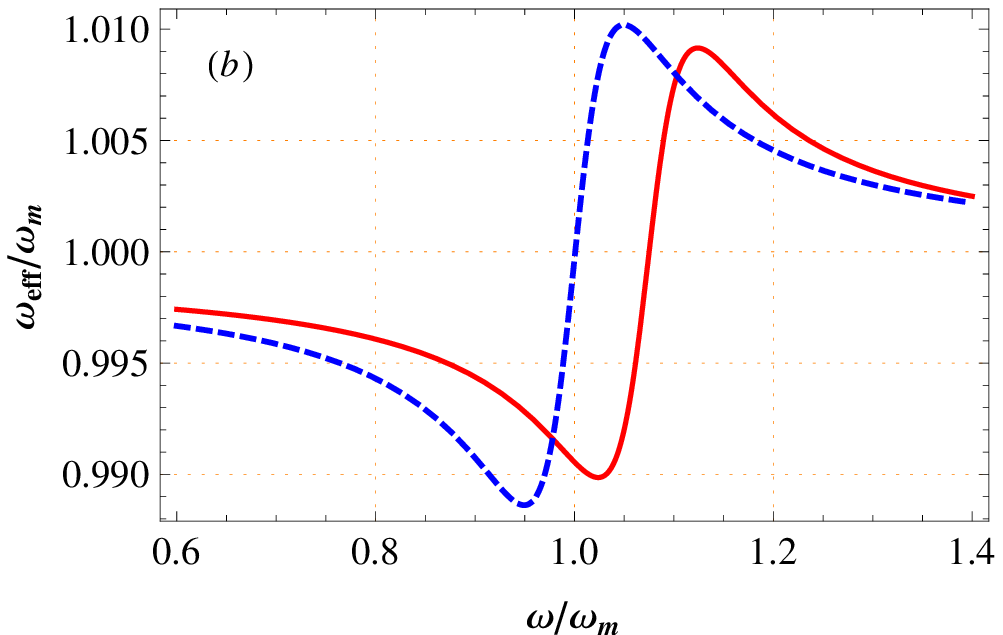}
\caption{
(Color online)The normalized  effective frequency  $\omega_{eff}/\omega_{m}$ versus the  $\omega/\omega_{m}$  (a) for $\eta=0.01$Hz and for different values of the  gain parameter:  $G=0$ (blue dashed curve), and  $G=10^6$Hz (red solid curve), (b) for $G=10^6$Hz and for   different values of anharmonicity parameter: $\eta=0$ (blue dashed curve), and $\eta=0.01$Hz (red solid curve). The parameters are   $\omega_{m}/2\pi=10$  MHz, $L=2$mm, $m=10$ ng, $Q= 5\times10^5 $ Hz, $P=6.9$ mW, $\kappa=0.01\omega_{m}$  , $\Delta=\omega_{m}$, and $\theta=\pi/2$ .
}
\label{fig:oeff}
\end{figure}

 The radiation-pressure force can   modify the dynamics of the mechanical oscillator and therefore the susceptibility of the oscillating mirror can be considered as the   susceptibility of an oscillator , with an effective frequency  and an effective  damping.These quantities contain important information about the quantum behavior of the system. The radiation presure-induced change in the frequency of the mechanical mode   is known as the optical spring effect \cite{spring} , and the  radiation presure-induced change in the damping rate  of the mirror shows how the optical field acts effectively as a viscous fluid
that  damps the mirror oscillation and cools its center-of mass motion\cite{rae,spring1}.
  In this section we investigate the effects of the  Kerr and the gain nonlinearities on the effective frequency $\omega_{eff}$ and the effective damping rate $\gamma_{eff}$ of the oscillating mirror. We can  find the mechanical susceptibility
of the  mirror according to the dependence of  the $\delta q (\omega)$ on the fluctuations in the total force on the
mirror $\delta F_{T}(\omega)$\cite{cohadon},
\begin{equation}
\delta q (\omega)=\chi (\omega)\delta F_{T}(\omega).
\end{equation} 
According to Eq.(\ref{deltaq}), $\delta F_{T}(\omega)$ consists of a radiation pressure and a Brownian motion term. Thus the mechanical susceptibility of the mirror $( \chi)$, its effective resonance frequency $( \omega_{eff})$, and its effective  damping rate $( \gamma_{eff})$ are, respectively, given by
 \begin{equation}
 \chi^{-1}=m(\omega_{eff}^{2}-\omega^{2})+i\omega\gamma_{eff},
 \end{equation}
 \begin{equation}\label{omegaeff}
 \omega_{eff}^{2} =\omega_{m}^{2}-\frac{\beta _{1}(\beta _{2}+\kappa^{2}-\omega^{2})}{|\beta _{2}+(-\kappa+i\omega)^{2}|^{2}},
 \end{equation}
 \begin{equation}\label{gammaeff}
 \gamma_{eff}=m(\gamma_{m}+\frac{2\beta _{1}\kappa}{|\beta _{2}+(-\kappa+i\omega)^{2}|^{2}} ).
 \end{equation}
In Figs.\ref{fig:oeff}(a) and \ref{fig:oeff}(b) we have plotted the normalized effective frequency of the oscillating mirror $\omega_{eff}/\omega_{m}$ versus the normalized frequency $\omega/\omega_{m}$  for $\Delta=\omega_{m}$ and for  different values of $G$ [Fig.(\ref{fig:oeff}a)] and different values of  $\eta$ [Fig.(\ref{fig:oeff}b)]. As   is seen, although the gain medium does not change the effective frequency of the moving mirror considerably , the energy shift of the cavity mode due to the  Kerr nonlinearity manifests itself in the  response frequency of the oscillating mirror. This means that the  effective coupling between the cavity and the mechanical modes (and therefore the cooling of the mirror) takes place when the effective detuning of the cavity $\Delta^{'}$ (but not $\Delta$) is chosen near the oscillating mirror frequency.
\begin{figure} [ht]
\centering
\includegraphics[width=3.5 in]{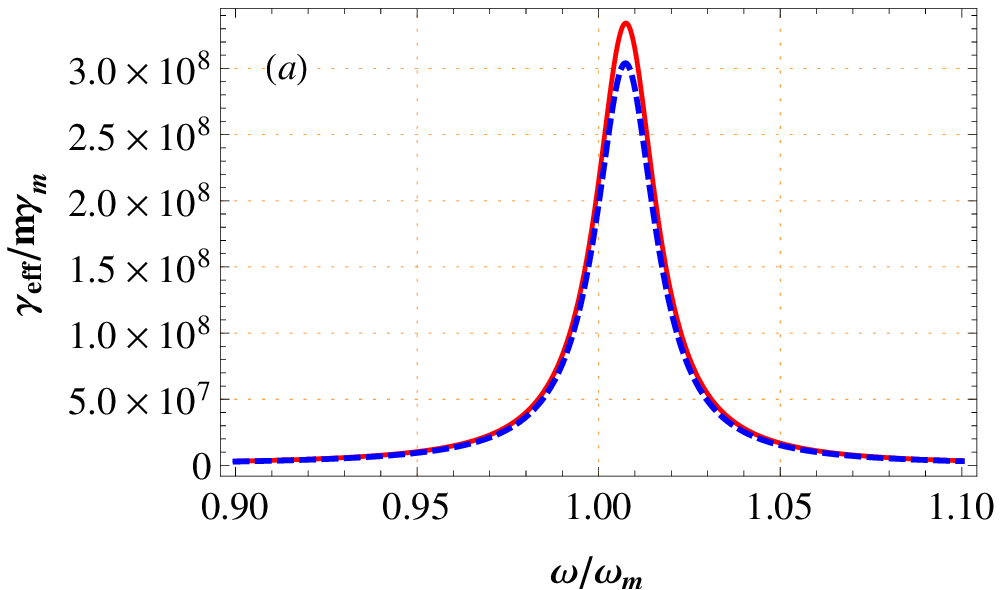}
\includegraphics[width=3.5 in]{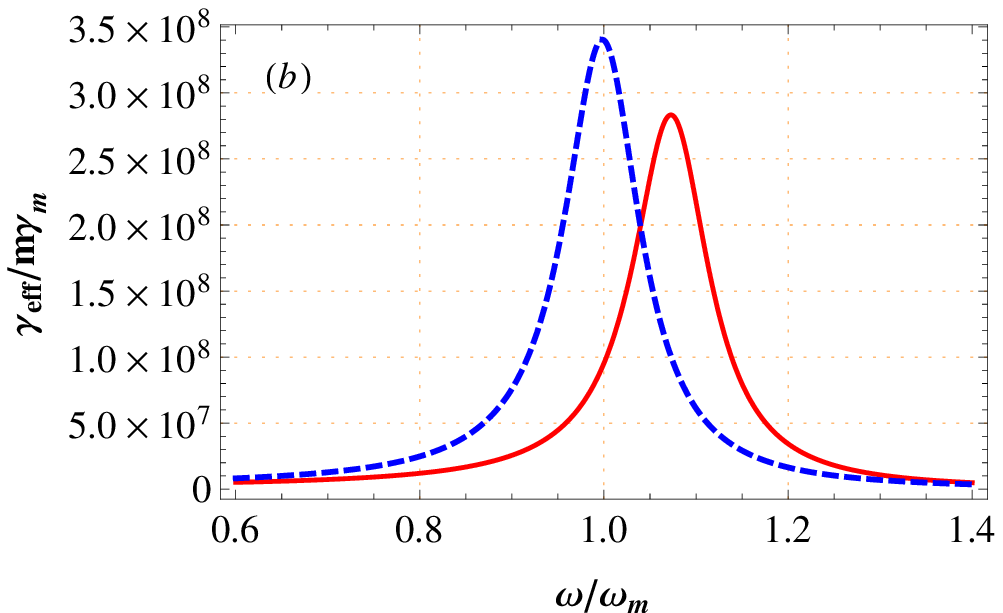}
\caption{
(Color online)The normalized  effective   damping rate of the oscillating mirror $\gamma_{eff}/m \gamma_{m}$ versus the  $\omega/\omega_{m}$ (a) for $\eta=0.01$Hz and for different values of gain parameter: $G=0$ (blue dashed curve), and  $G=10^6$Hz (red solid curve), (b) for $G=10^6$Hz and for   different values of anharmonicity parameter: $\eta=0$ (blue dashed curve), and $\eta=0.01$Hz (red solid curve). The parameters are   $\omega_{m}/2\pi=10$  MHz, $L=2$mm, $m=10$ ng, $Q= 5\times10^5 $ Hz, $P=6.9$ mW, $\kappa=0.01\omega_{m}$, $\Delta=\omega_{m}$, and $\theta=\pi/2$ .
}
\label{fig:dmpeff}
\end{figure}
In Figs.\ref{fig:dmpeff}(a) and  \ref{fig:dmpeff}(b) we have plotted the normalized effective damping rate of the oscillating mirror $\gamma_{eff}/m \gamma_{m}$ versus the normalized  frequency $\omega/\omega_{m}$ for $\Delta=\omega_{m}$ and for different values of $G$ [Fig.\ref{fig:dmpeff}(a)] and different values of $\eta$ [Fig.\ref{fig:dmpeff}(b)]  . As  is seen, by increasing  the gain parameter the effective damping rate of the mirror motion increases  while increasing the anharmonicity parameter not only leads to the  frequency shift but also decreases the effective damping rate.

\subsection{Effects of the Kerr-down conversion nonlinearity on the effective tempreture of the mirror}
  Here we examine the effects of the  Kerr and the gain nonlinearities on the back-action ground-state cooling of the oscillating mirror. In order to investigate the cooling of the mirror it is sufficient to consider the mean number of quanta of the vibrational    excitation of the mirror as defined by\cite{bhat-cooling}
 \begin{equation}
 n_{m}=\frac{k_{B}T_{eff}}{\hbar \omega_{eff}}=\frac{k_{B}T}{\hbar \omega_{m}}\dfrac{\gamma_{m}}{\gamma_{eff}}(\dfrac{\omega_{m}}{\omega_{eff}})^{3},
  \end{equation}
where we expand the effective frequency and effective damping rate of the oscillating mirror in a Taylor series around $\omega=\omega_{m}$
 and keep only the leading terms in the respective expansions
\begin{eqnarray}
\omega_{eff}(\omega) \sim\omega_{eff}(\omega_{m})\equiv\omega_{eff}\nonumber, \\
\gamma_{eff}(\omega)\sim \gamma_{eff}(\omega_{m})\equiv\gamma_{eff}.
\end{eqnarray}
The ground-state cooling is approached if $n_{m} < 1$. 
In  Figs.\ref{fig:n,teff}(a) and \ref{fig:n,teff}(b) we have plotted, respectively,  $n_{m}$ and $T_{eff}$ for the initial equilibrium temperature $T=0.4$K  versus $\Delta/\omega_{m}$ and   for  different values of $G$ and $\eta$. It is evident that as $G$ increases    the system  cools down to the ground-state ($n_{m}<1$) while as $\eta$ increases the effective temperature of the system increases . The heating of the system in the presence of the Kerr medium is due to the energy shift of the cavity  which  weakens  the field-mirror coupling and reduces the number of intracavity photons because of the photon-photon repulsion mechanism.
\begin{figure} [ht]
\centering
\includegraphics[width=3.5 in]{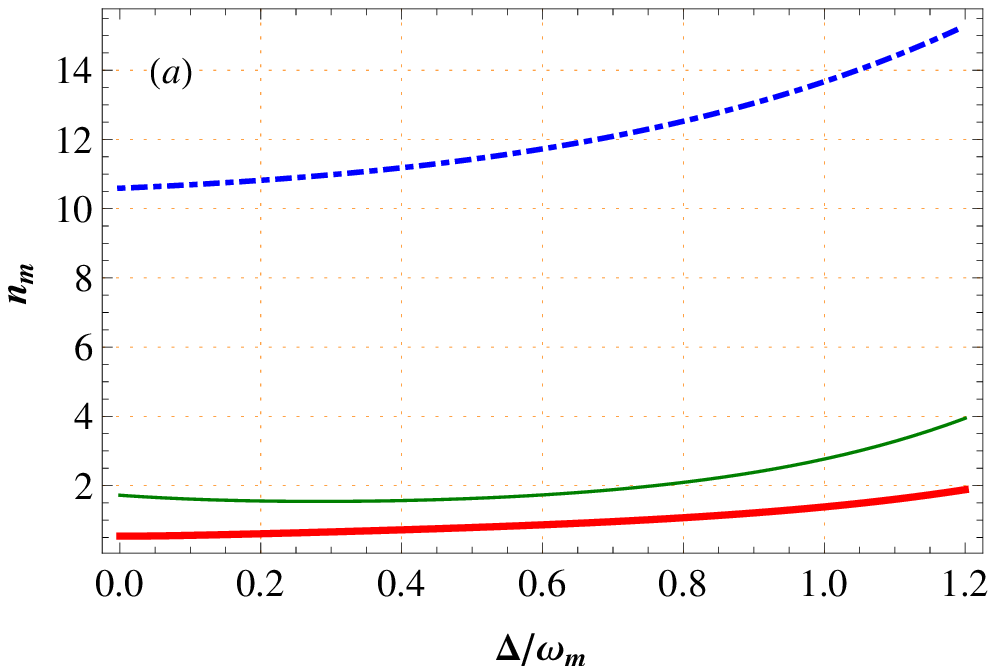}
\includegraphics[width=3.5 in]{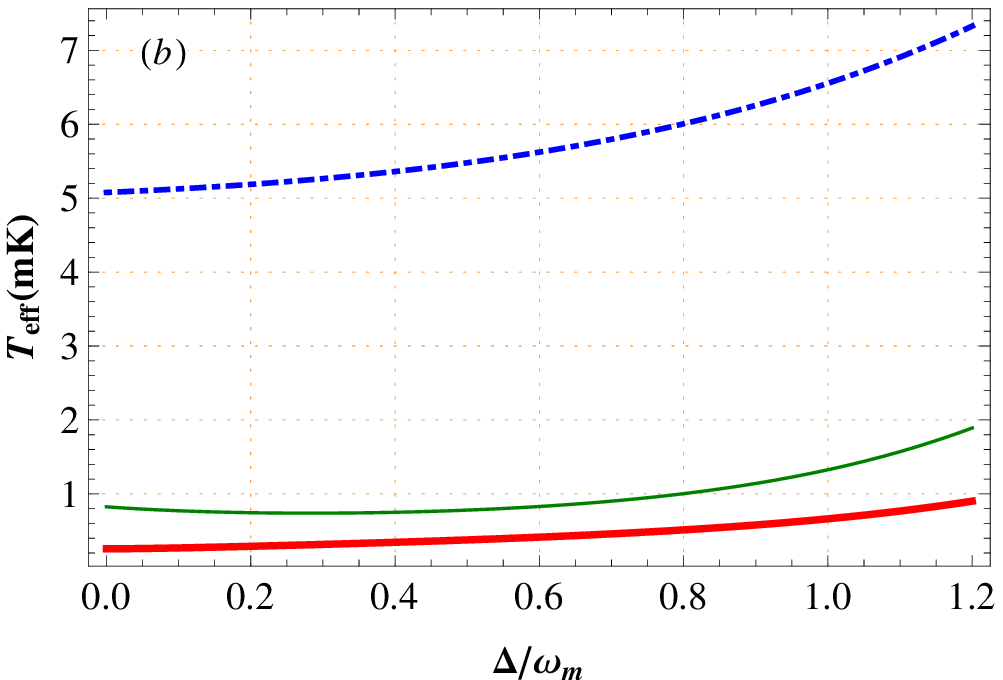}
\caption{
(Color online) Plots of (a) the  mean number of vibrational quanta $n_{m}$ and (b) the effective temperature $T_{eff}$ for $G=2\times10^7$Hz,$\eta=0.05$Hz (blue dotted-dashed curve),   $G=2\times10^7$Hz,$\eta=0.01$Hz (red thick curve) and $G=0,\eta=0.01$Hz  (green thin curve). The parameters are   $\omega_{m}/2\pi=10$  MHz, $L=10$mm,  $m=10$ ng, $Q= 5\times10^5 $ Hz, $P=6.9$ mW, $\kappa=0.7\omega_{m}$,  and $\theta=\pi/2$ .
}
\label{fig:n,teff}
\end{figure}

 \section{Effects of the Kerr-down conversion nonlinearity on the intensity and quadrature noise spectra of the transmitted field}\label{field}
   
\begin{figure} [ht]
\centering
\includegraphics[width=3.5 in]{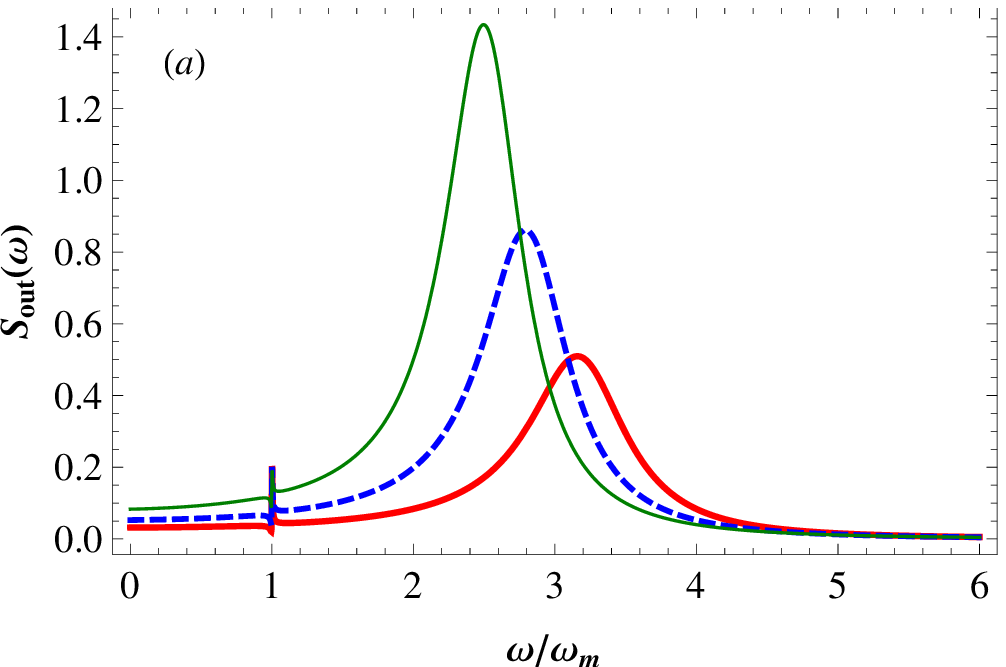}
\includegraphics[width=3.5 in]{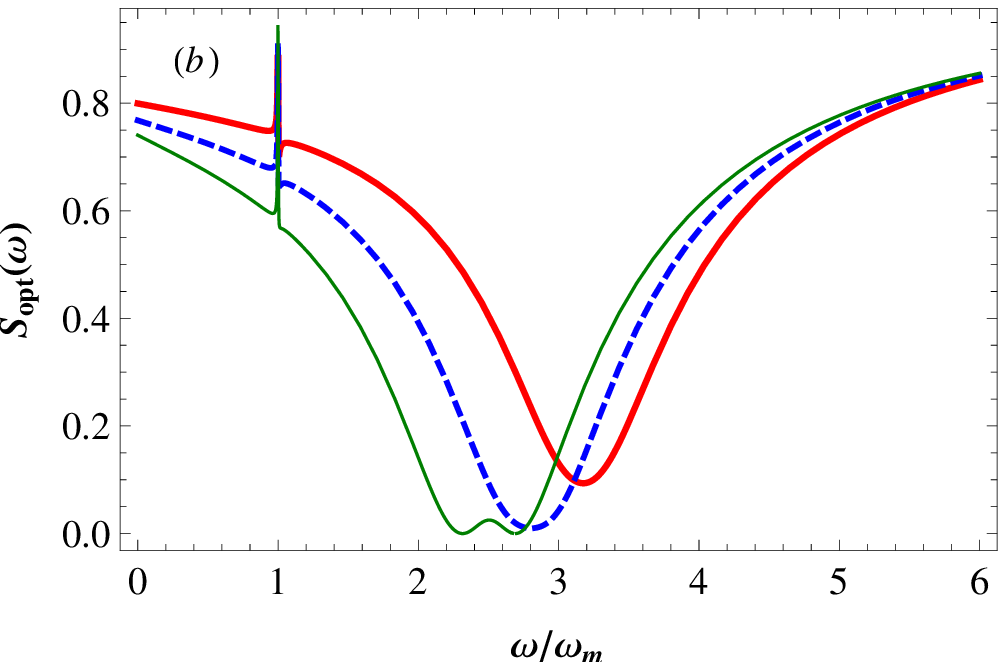}
\caption{
(Color online)(a) The  intensity spectrum and (b)the squeezing spectrum of the transmitted field versus the normalized frequency $\omega/\omega_{m}$ for  $G=1\times10^{7}$ Hz (red thick curve), $G=2\times10^7$ Hz(blue dashed  curve),  and  $G=3\times10^7$ Hz(green thin curve). The parameters are  $\eta=0.01$Hz,  $\omega_{m}/2\pi=10$  MHz, $L=10$mm,  $m=10$ ng, $Q= 5\times10^5 $ Hz, $P=6.9$ mW, $\kappa=0.7\omega_{m}$,  and $\theta=\pi/2$ .
}
\label{fig:fi-g}
\end{figure}

\begin{figure} [ht]
\centering
\includegraphics[width=3.5 in]{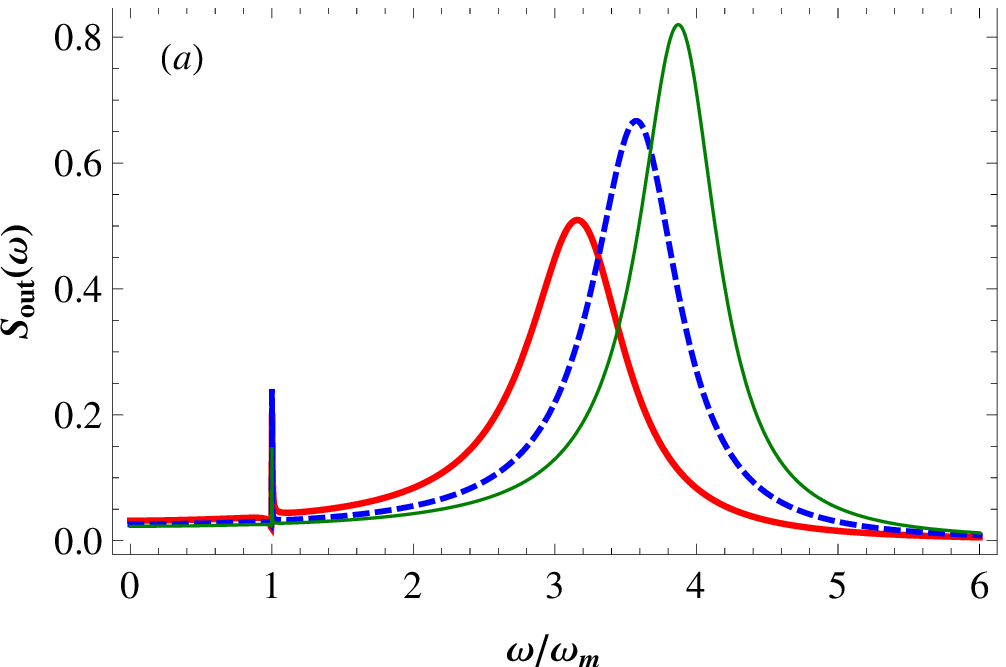}
\includegraphics[width=3.5 in]{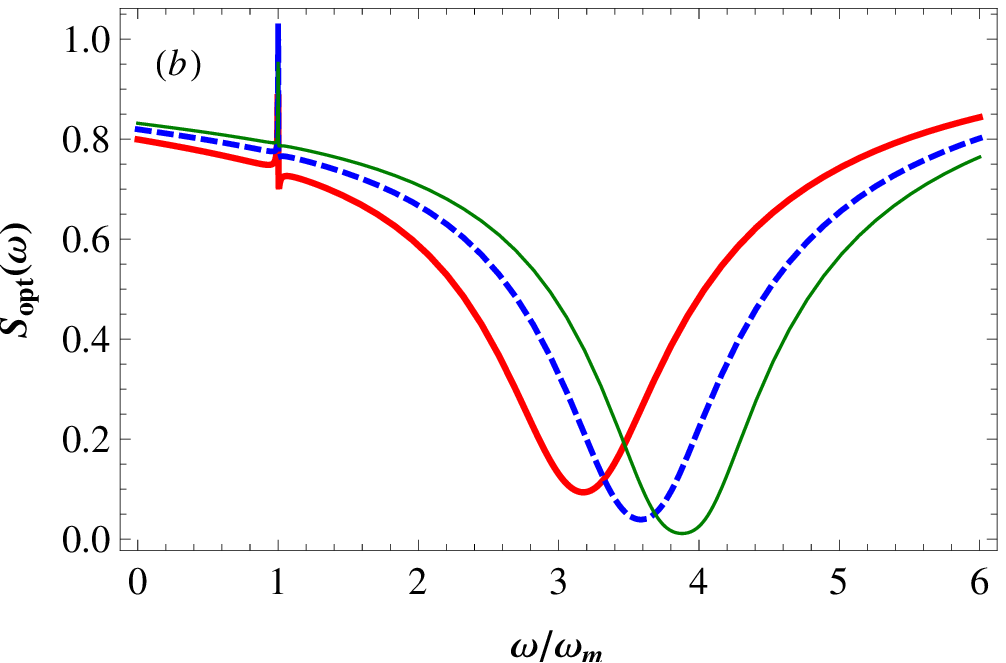}
\caption{
(Color online)(a) The  intensity spectrum and (b)the squeezing spectrum of the transmitted field versus the normalized frequency $\omega/\omega_{m}$ for $\eta=0.01$ Hz (red thick curve), $\eta=0.03$ Hz(blue dashed  curve),  and  $\eta=0.05$ Hz(green thin curve). The parameters are  $G=10^7$Hz, $\omega_{m}/2\pi=10$  MHz, $L=10$mm,  $m=10$ ng, $Q= 5\times10^5 $ Hz, $P=6.9$ mW, $\kappa=0.7\omega_{m}$,  and $\theta=\pi/2$ .
}
\label{fig:fi-eta}
\end{figure}

 The intensity spectrum of the transmitted field is given by the Fourier transform of the two time correlation functions $\langle\delta a^{\dagger}_{out}(t+\tau)\delta a_{out}(t)\rangle$:
 \begin{equation}
 S(\omega)=\int_{-\infty}^{\infty} d\tau\langle\delta a^{\dagger}_{out}(t+\tau)\delta a_{out}(t) \rangle e^{-i\omega \tau}.
 \end{equation}
 Solving the matrix equation (\ref{matrixform})  we obtain
the solution for the cavity-field fluctuation operator $\delta a $ and then using the input-output relation $a_{out}=\sqrt{2\kappa}a-a_{in}$, we obtain the fluctuations  of the output field to be 
\begin{equation}
\delta a_{out}(\omega)=V_{1}(\omega)\xi(\omega)+V_{2}(\omega)\delta a_{in}(\omega)+V_{3}(\omega)\delta a^{\dagger}_{in}(\omega),
\label{aout}
\end{equation}
and $\delta a^{\dagger}_{out}(\omega)=[\delta a_{out}(-\omega)]^{\dagger}$,
where
 \begin{eqnarray}
 V_{1}(\omega)&=&\sqrt{2\kappa} \dfrac{i g_{m} }{ M d(\omega)} 
 \lbrace a_{s}(\kappa -i\Delta^{'}-i\omega)-2 G a_{s}^{*}e^{i\theta}\rbrace, \nonumber \\
 V_{2}(\omega)&=&\dfrac{2\kappa}{ d(\omega)}\lbrace 
\dfrac{i\hbar g_{m}^{2}a_{s}^{2}}{M}+(\omega_{m}^{2}-\omega^{2}+i\omega \gamma_{m}) \nonumber\\
 && (i \delta_{1}+\Gamma)\rbrace -1, \nonumber \\
 V_{3}(\omega)&=&\dfrac{2\kappa}{ d(\omega)}\lbrace\dfrac{i\hbar g_{m}^{2}\vert a_{s}\vert^{2}}{M}+(\omega_{m}^{2}-\omega^{2}+i\omega \gamma_{m})\nonumber\\
 && (\kappa+i(\omega+\Delta_{1}))\rbrace, \label{a-coeff}
 \end{eqnarray} 
 In  Eq.(\ref{aout}) the first term stems from the mechanical oscillator thermal noise, while the  other two terms are from the input vacuum noise. 
 Making use of the correlation peroperties for the  noise forces and  Eqs.(\ref{aout}) and (\ref {a-coeff}), we obtain the output field spectrum as 
 \begin{equation}
 S(\omega)=\vert V _{3}(-\omega)\vert ^{2} +m \hbar \gamma_{m}\omega[1+\coth(\frac{\hbar\omega}{2k_{B}T})]\vert V _{1}(-\omega)\vert ^{2}.
 \end{equation}
The quadrature noise spectrum of the transmitted filed is given by\cite{eluch}
 \begin{eqnarray}
 S_{\varphi}(\omega)&=&\int_{-\infty}^{\infty} d\tau \langle \delta x^{out}_{\varphi}(t+\tau)  \delta x^{out}_{\varphi} (\tau)  \rangle_{ss} e^{-i\omega \tau}\nonumber\\
 &=&  \langle\delta x^{out}_{\varphi}(\omega)  \delta x^{out}_{\varphi}(\omega)\rangle ,
 \end{eqnarray}
 where $\delta x^{out}_{\varphi}(\omega)=e^{-i\varphi}\delta a_{out}(\omega)+e^{i\varphi}\delta a^{\dagger}_{out}(\omega) $ is the Fourier trnasform of the output quadrature, with $\varphi$ as its externally controllable   phase angle  which is experimentally  measurable in a homodyne detection scheme\cite{homodyne}.For the  system under consideration  quadrature noise spectrum is given by
 \begin{eqnarray}
  S_{\varphi}(\omega)&=&e^{-2i\varphi} C_{aa}^{out} (\omega)+e^{2i\varphi} C_{aa}^{out*} (\omega)\nonumber\\
  &+&
  C_{a^{\dagger}a}^{out} (\omega)+C_{aa^{\dagger}}^{out}  (\omega),
  \label{squeezing}
 \end{eqnarray}
 where
  \begin{eqnarray}
C_{aa}^{out} &=&\lbrace m \hbar \gamma_{m}\omega[1+\coth(\frac{\hbar\omega}{2k_{B}T})] V_{1}(\omega)V_{1}(-\omega))\nonumber\\
  &+&V_{2}(\omega)V_{3}(-\omega)\rbrace,\\
  C_{aa^{\dagger} }^{out} &=&\lbrace m \hbar \gamma_{m}\omega[1+\coth(\frac{\hbar\omega}{2k_{B}T})]\vert V_{1}(\omega)\vert ^{2}\nonumber\\
  &+&\vert V_{2}(\omega)\vert ^{2}\rbrace,\\
  C_{a^{\dagger}a}^{out} &=&\lbrace m \hbar \gamma_{m}\omega[1+\coth(\frac{\hbar\omega}{2k_{B}T})]\vert V_{1}(-\omega)\vert ^{2}\nonumber\\
  &+&\vert V_{3}(-\omega)\vert ^{2} \rbrace.
 \end{eqnarray}
 We  define the optimum quadrature squeezing $S_{opt}(\omega)$ by choosing $\varphi(\omega)$ in such a way that $dS_{\varphi}(\omega)/d\varphi=0$. This yields
 \begin{equation}
e ^{2i \varphi_{opt}}=-\dfrac{C_{aa}^{out}}{\vert C_{aa}^{out}\vert}.
 \end{equation}
 Then, substituting back into Eq.(\ref{squeezing}), we obtain the optimized squeezing  spectrum as
 \begin{eqnarray}
  S_{opt}(\omega)&=& -2 \vert  C_{aa}^{out}\vert +C_{a^{\dagger}a}^{out} (\omega)+C_{aa^{\dagger}}^{out}  (\omega).  
 \end{eqnarray}
 In Figs.\ref{fig:fi-g}(a) and \ref{fig:fi-g}(b) we have  plotted, respectively, the   intensity and the optimized  squeezing spectra of the transmitted field  for  different values of the gain nonlinearity  versus the normalized response frequency $\omega/\omega_{m}$. As can be seen  increasing the gain parameter $G$  causes the intensity   and the  degree of squeezing of the transmitted field to increase. The shift in the intensity spectrum is due to  the energy shift of the cavity in the presence of the Kerr medium.
 In Figs.\ref{fig:fi-eta}(a) and \ref{fig:fi-eta}(b) we have plotted, respectively, the  intensity and the optimized squeezing spectra  of the transmitted field  for  different values of  the anharmonicity parameter $\eta$ versus the normalized response frequency $\omega/\omega_{m}$. It can be seen  as $\eta$ increases the intensity   and the  degree of squeezing of the transmitted field  increases. 
  \section{Entanglement peroperties of the system}\label{ent}
It has been shown previously \cite{ent} that within  each optomechanical cavity  the  photon-phonon entanglement can be generated by means of radiation pressure . Here we are interested in the generation  of stationary entanglement between the movable mirror and the  cavity field in the presence of Kerr-down conversion medium. Since our system is in a Gaussian state and the linear nature of  Eq.(\ref{matrixform}) preserves the Gaussian nature of the initial state of the system,  we can use the logarithmic negativity measure \cite{measure} to quantify the entanglement. When the stability conditions of Eq.(\ref{routh1}) are fulfilled, we can solve   Eq.(\ref{matrixform}) for the  $4\times4$ stationary correlation  correlation matrix (CM) $V$
\begin{equation}\label{ly}
M V+V M^{T}=-D,
\end{equation}
where the elements of the correlation matrix $V$ are defined as $  V_{ij} = \langle  u_{i}u_{j}+u_{j}u_{i}\rangle/2$ and $D=Diag[0,m\hbar\gamma_{m}\omega_{m}(2\bar{n} +1),\kappa,\kappa]$   is the diagonal diffusion matrix in which we used the following apporximation 
 \begin{equation}
 \omega \coth(\frac{\hbar\omega}{2 k_{B}T})\simeq \omega_{m}\frac{2 k_{B}T}{\hbar\omega_{m}}\simeq \omega_{m}(2\bar{n}+1),
 \end{equation}
 where $\bar{n}=[e^{\hbar\omega_{m}/k_{b}T}-1]^{-1}$.
 Then the photon-phonon entanglement   can be quantified by the logarithmic negativity $E_{N}$ as 
 \begin{equation}
E_N=\mathrm{max}[0,-\mathrm{ln} 2 \eta^-],,
 \end{equation}
   where  $\eta^{-} \equiv2^{-1/2}\left[\Sigma(V)-\sqrt{\Sigma(V)^2-4 \mathrm{det} V}\right]^{1/2}$, is the lowest symplectic eigenvalue of the partialtranspose of the $4 \times 4$ CM, $V$, with $\Sigma(V)=\mathrm{det} V_{A} +\mathrm{det}V_{B} -2\mathrm{det} V_{C}$, and we used the $2\times 2$ block form of the CM 
   \begin{equation}
 \left( \begin{matrix}
 V_{A} &V_{C}  \\ 
V_{C}^{T} & V_{B} 
\end{matrix} \right),
  \end{equation}   
 with $V_{A}$ associated to the oscillating mirror, $V_{B}$  to the cavity mode, and $V_{C}$ describing the optomechanical correlations.
  \begin{figure}[ht]
\centering
\includegraphics[width=3in]{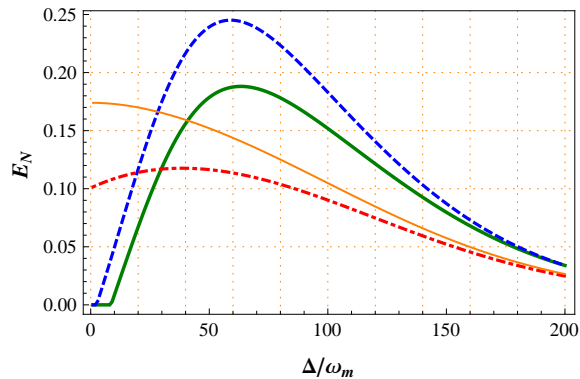}
\caption{
(Color online) The logarithmic negativity $E_{N}$  versus the normalized detuning $\Delta/\omega_{m}$ for  $(\eta=0, G=0)$ (green thick curve), $(\eta=0.01\mathrm{Hz}, G=0)$(red dashed-dotted  curve),  $(\eta=0 , G=6\times10^{6}\mathrm{Hz})$(blue dashed  curve) and  $(\eta=0.01\mathrm{Hz} , G=6\times10^{6}\mathrm{Hz})$(orange thin curve). The parameters are $\omega_{m}/2\pi=10$  MHz, $L=1$mm, $m=10$ ng $Q= 5\times10^5 $ Hz,  $P=15$ mW,  $\kappa=0.9\omega_{m}$,  and $\theta=\pi/2$.}
\label{fig:figent}
\end{figure}  
 \begin{figure}[hb]
\centering
\includegraphics[width=3in]{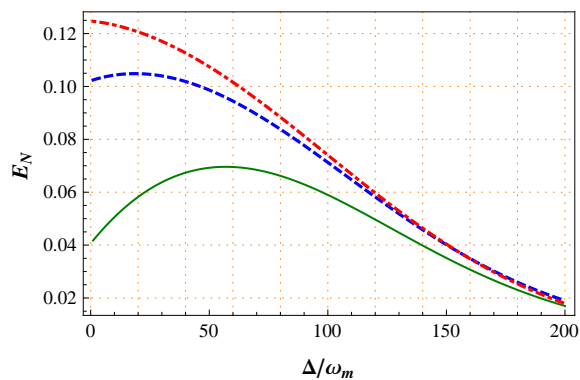}
\caption{
(Color online) The logarithmic negativity $E_{N}$  versus the normalized detuning $\Delta/\omega_{m}$ for  $ G=10^{3}$ Hz(green solid curve), $ G=10^{6}$ Hz(red dashed-dotted curve), and $ G=10^{7}$Hz(blue dashed  curve) . The parameters are  $\eta=0.01$ Hz, $\omega_{m}/2\pi=10$  MHz, $L=1$mm,  $m=10$ ng, $Q= 5\times10^5 $ Hz,  $P=15$ mW,  $\kappa=0.9\omega_{m}$,  and $\theta=\pi/2$. }
\label{fig:figent2}
\end{figure}  
 In Fig.(\ref{fig:figent}) we have plotted $E_{N}$  versus the normalized detuning $\Delta/\omega_{m}$ for a bare cavity, a cavity that contains only the Kerr medium, a cavity that contains only the gain medium and a cavity that contains both the  Kerr and gain medium.
 We see that, as expected, while the Kerr medium reduces the degree of phonon-photon entanglement, the gain medium increases it considerably. This result arises from the fact that the  radiation pressure, which is responsible for the phonon-photon entanglement,  increases considerably in the presence of the gain nonlinearity while decreases in the presence of the Kerr nonlinearity. Therefore similarly to what happens for the mirror cooling process  there is a  competition between the gain and Kerr  nonlinearities  for increasing and decreasing quantum effects, respectively.
 In Fig.\ref{fig:figent2} we have plotted $E_{N}$  versus the normalized detuning for different values of the  gain nonlinearity. It can be seen that as $G$ increases the degree of entanglement increases. It should be noted that  we cannot increase $G$ arbitrarily because of the stability condition of the system. 
\section{Summary and Conclusions}\label{sum}
  In this paper we  considered  the canonical optomechanical cavity composed of a partiallly transmitting mirror and one perfectly reflecting movable mirror containing a Kerr-down conversion nonlinear medium. For this system we investigated the influence of the two nonlinearities on the dynamics of the oscillating mirror, the  intensity and squeezing spectra of the  transmitted field and the steady-state mirror-field  entanglement.
  For  the dynamics of the oscillating mirror we have found  that  in the  peresence of the two nonlinearities, there is a competition between the two  quantum processes; the splitting of the  pump photons in two degenerate subharmonic photons  as well as increasing  the  intracavity photon number  due to  the nonlinear gain medium, and  shifting  the energy level of the cavity proportional to the anharmonicity parameter $\eta$ which blocks the entrance of a large number of photons into the cavity. This competition can lead to enhancement of the radiation pressure contribution to the normal modes of the cavity and NMS below the threshold $g_{0}a_{s}\lesssim \kappa$.  Also, it was demonstrated that in the presence of the  Kerr-down conversion nonlinearity the ground-state cooling of the mechanical mirror is possible. Although with increasing the anharmonicity parameter $\eta$ the  vibrational excitation and effective temperature of the mirror increases,  the enhancement of the gain medium can compensate the increasing of the  phonon number to some extent.
  In the investigation of the intensity spectrum and quadrature squeezing of the transmitted field, it was shown that the range of response frequency of the field undergoes a shift because of anharmonicity parameter $\eta$ but in this range of the frequency both the  Kerr and gain media increase the  degree of squeezing.
Finally, we investigated the influence of the   Kerr-down conversion on the degree of stationary  entanglement between the cavity and the mechanical modes. It was shown that  the Kerr nonlinearity reduces the degree of entanglement considerably, while the gain medium increases it.   
 \section*{Acknowledgement}
The authors wish to thank The Office of Graduate
Studies of The University of Isfahan for their support.
\bibliographystyle{apsrev4-1}

\end{document}